\documentclass[a4paper]{jpconf}
\usepackage{graphicx}
\begin{document}
\title{Dynamics aspect of subbarrier fusion reaction in light heavy ion systems}

\author{M. Huang$^1$, F. Zhou$^2$, R. Wada$^{1,*}$, X. Liu$^{1,3}$, W. Lin$^{1,3}$, M. Zhao$^{1,3}$, J. Wang$^1$, Z. Chen$^1$, C. Ma$^4$, Y. Yang$^{1,3}$, Q. Wang$^1$, J. Ma$^1$, J. Han$^1$, P. Ma$^1$, S. Jin$^{1,3}$, Z. Bai$^{1,3}$, Q. Hu$^{1,3}$, L. Jin$^{1,3}$, J. Chen$^{1,3}$, Y. Su$^{1,3}$ and Y. Li$^{1,3}$}
\address{$^1$ Institute of Modern Physics, Chinese Academy of Sciences, Lanzhou, 730000,China.}
\address{$^2$ School of Nuclear Science and Technology, University of Science and Technology of China, Hefei, Anhui 230026, China.}
\address{$^3$ Graduate University of Chinese Academy of Sciences, Beijing, 100049, China.}
\address{$^4$ Department of Physics, Henan Normal University, Xinxiang, 453007, China.}
\ead{M. Huang, huangmeirong@impcas.ac.cn}
\ead{*Corresponding author: R. Wada, wada@comp.tamu.edu}

\begin{abstract}
Subbarrier fusion of the $^{7}Li + ^{12}$C reaction is studied using an antisymmetrized molecular dynamics model (AMD) with an after burner, GEMINI. In AMD, $^{7}Li$ shows an $\alpha + t$ structure at its ground state and it is significantly deformed. Simulations are made near the Coulomb barrier energies, i.e., E$_{cm} = 3 - 8 MeV$. The total fusion cross section of the AMD + GEMINI calculations as a function of incident energy is compared to the experimental results and both are in good agreement at E$_{cm} > 3 MeV$. The cross section for the different residue channels of the AMD + GEMINI at $E_{cm} = 5 MeV$ is also compared to the experimental results.
\end{abstract}

\section{Introduction}

Availability of radioactive beam facilities has stimulated theoretical and experimental interest in the structure of nuclei far from the stability line. Nuclear fusion reactions near the Coulomb barrier are strongly affected by the structure of the interacting nuclei, especially with weakly bound neutrons~\cite{Canto06}. Some theoretical calculations predict that the fusion cross section is enhanced over well-bound nuclei because of the larger spatial extent of halo nucleons~\cite{Takigawa93}. On the other hand halo nuclei can easily break up in the field of the other nucleus, due to their low binding energies, before the two nuclei come close enough to fuse and carry away available energy. Early calculations of Hussein {\it et al.}~\cite{Hussein92} indicate that the actual fusion cross section decreases significantly at all energies. However recent couple channel calculations of Hagino {\it et al.}~\cite{Hagino00} have concluded that the fusion cross section increases below the Coulomb barrier because of the neutron halo whereas it decreases above the barrier because of weak coupling of the halo nucleons. Experimentally this is still a hot debate because of experimental difficulties. Another interest we propose here is the influence of the cluster structure in the fusion mechanism. The cluster structures of light nuclei have been studied theoretically from stable to those far away from the stability line. Recent calculations, using an antisymmetrized molecular dynamics model(AMD), indicate that light nuclei exhibit variety of distinct cluster structures ~\cite{Kaneda_En'yo95,Kaneda_En'yo95_1,Kaneda_En'yo99,Furutachi09}. The cluster structures are predicted even for nuclei with $Z \sim N$ of Li and Be~\cite{Kaneda_En'yo95_1} (where Z and N are the charge and neutron number in a nucleus, respectively). When nuclei with a well-developed cluster structure are involved in fusion reactions near the barrier, it will be reflected on the fusion cross section, especially in the variety of the exit channel distribution or in the fusion residue distribution. The cluster structure effect may be enhanced in the fusion reaction between light systems. In these systems the Coulomb interaction becomes small and the proximity effect between the two nuclei will be enhanced and therefore the structure of the projectile and/or target may reflect the fusion cross section directly, especially as an enhancement of particular incomplete fusion channels. In this paper we present the calculated results in the study of the fusion reactions of the $^7Li + ^{12}C$ system near the Coulomb barrier using AMD simulations.

\section{AMD simulations}

The initial nuclei of $^7Li$ and $^{12}C$ were produced by the AMD code of Ono {\it et al.} in refs.~\cite{Ono02}, using the Gogny interaction. The binding energy and root mean square radius of these initial nuclei are compared to the experimental values in Table 1. All calculated values are in good agreement to those of the experimental values, except for the root-mean square radius of $^7Li$ in which the calculated value is about $20\%$ larger than that of the experiment. The more sophisticated calculation in ref.~\cite{Kaneda_En'yo95_1} of the experimental root mean square radius of $^7Li$ is also well reproduced with a distinct $\alpha + t$ structure.

\begin{table}[h]
\caption{\label{tabone}initial nuclei.}
\begin{center}
\lineup
\begin{tabular}{*{5}{l}}
\br
$\0\0$nucleus&AMD&AMD&$\m$Exp.&$\m$Exp.\cr
$\0\0$&Binding&rms (fm)&$\m$Binding&$\m$rms (fm)\cr
$\0\0$&energy (MeV)&&$\m$energy (MeV)&$\m$\cr
\mr
\0\0$^{7}$Li(t,$\alpha$)&40.00 &3.02&$\m$39.24&$\m$2.43\cr
\0\0$^{12}$C(3$\alpha$)&92.24 &2.53&$\m$92.16&$\m$2.47\cr
\br
\end{tabular}
\end{center}
\end{table}

\begin{figure}[h]
\begin{center}
\begin{minipage}{14pc}
\includegraphics[width=14pc]{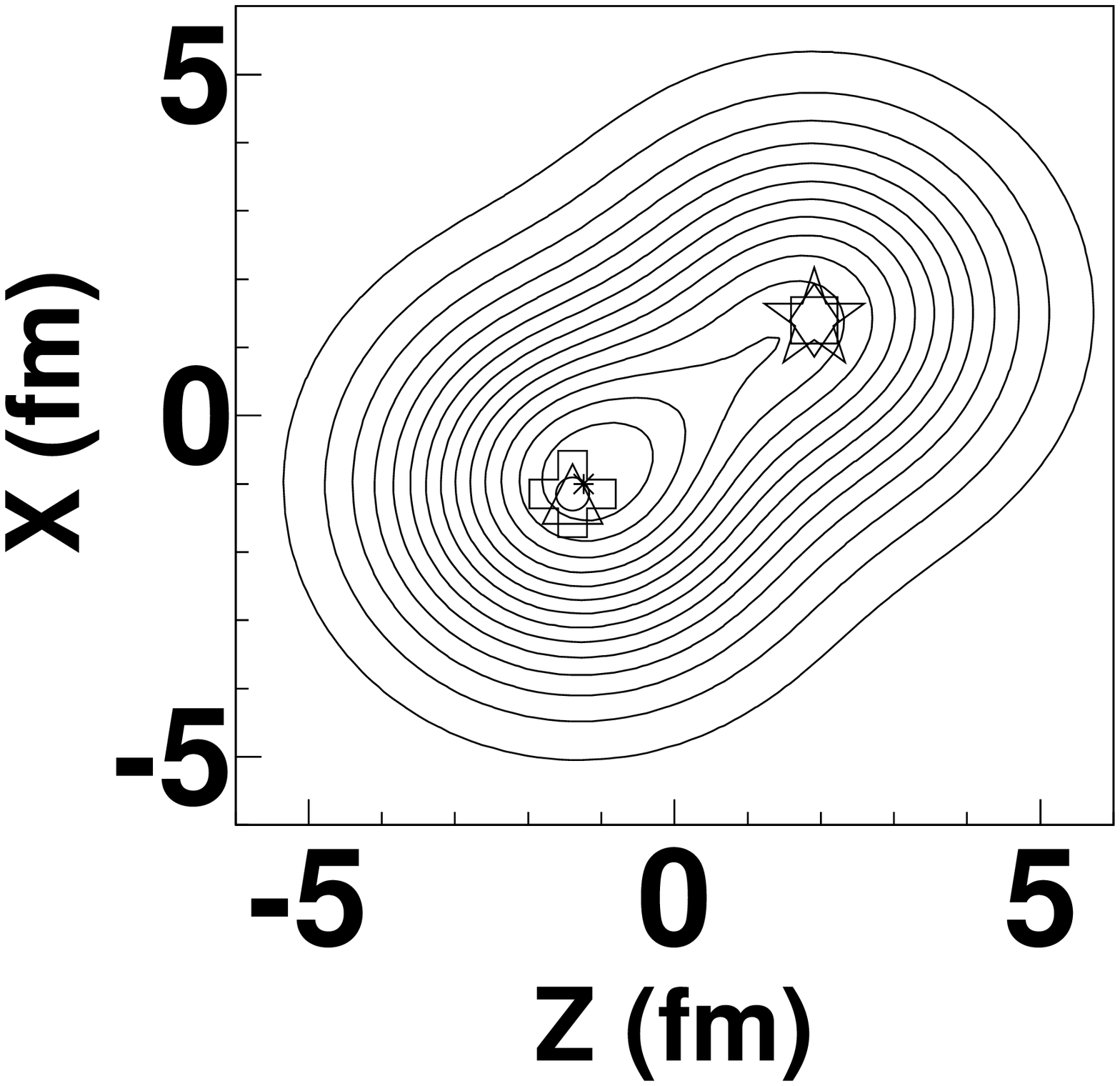}
\caption{\label{fig1a}$^{7}Li$ initial nucleus with $\alpha$ + t structure.}
\end{minipage}\hspace{2pc}%
\begin{minipage}{14pc}
\includegraphics[width=14pc]{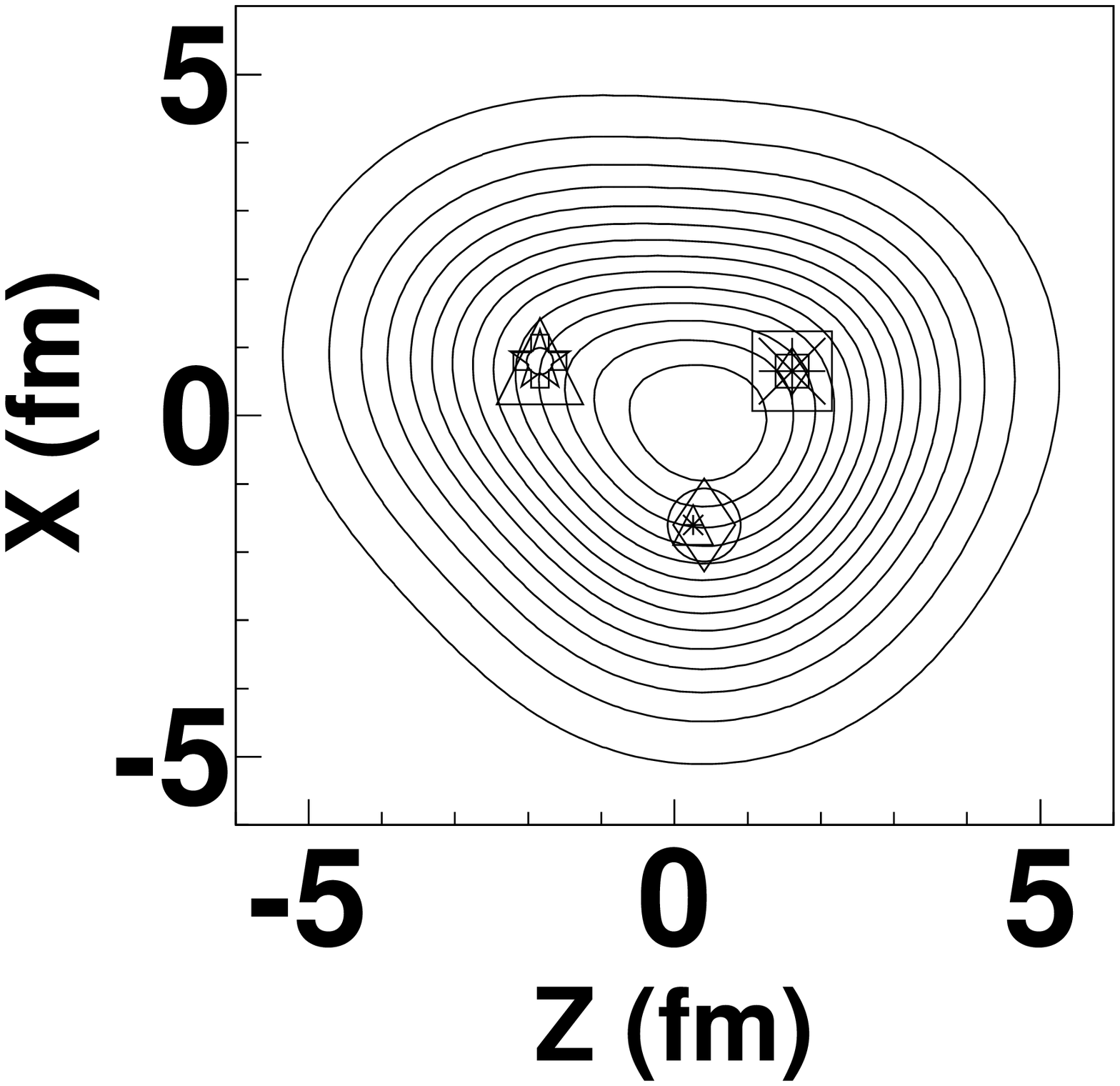}
\caption{\label{fig1b}$^{12}C$ initial nucleus with 3$\alpha$ structure.}
\end{minipage}
\end{center}
\end{figure}

The density plot of these nuclei are also shown in Fig.\ref{fig1a} and Fig.\ref{fig1b}. Symbols indicate the location of all nucleons. One can see in both figures that nucleons are well clusterized in space. In the $^7Li$ case, two clusters are observed, the larger one corresponds to an $\alpha$ and the other to a triton, and the nucleus is very deformed. In $^{12}C$, $3 \alpha$ clusters are observed, but the nucleus is compact and much more spherical.

\begin{figure}[h]
\begin{center}
\includegraphics[scale=0.45]{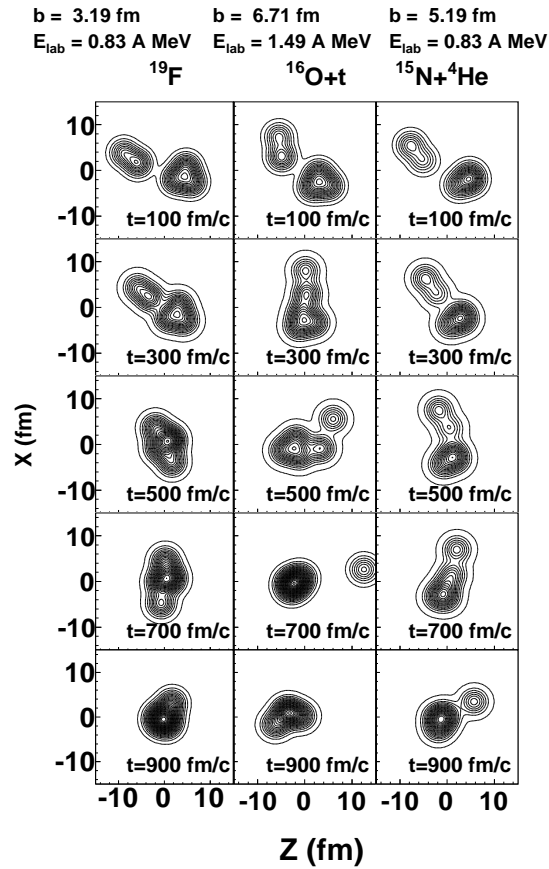}
\caption{
Time evolution of the 2D density plots for typical fusion reactions. Impact parameter, incident energy and reaction product at the bottom of the simulations are indicated on the top of each figure. The density plot is made by projecting that of all nucleons on the X-Z plane. The contour lines are plotted on a linear scale.
}
\label{fig:fig2}
\end{center}			
\end{figure}

Using these initial nuclei, $^7Li + ^{12}C$ reactions were simulated at center of mass energies between 3 to 8 MeV. Calculations were performed in the impact parameter range, b, from 0 to $7 fm$. In $ b > 7  fm$, no fusion reactions are observed. In Fig.\ref{fig:fig2} the time evolution of the density distributions is shown as typical examples of the complete and incomplete fusion reactions. On the left panel, a complete fusion reaction is observed. In the middle, only the $\alpha$ particle is transferred into the $^{12}C$ nucleus and triton is escaped as a spectator. On the right panel, only triton is absorbed and the $\alpha$ particle becomes a spectator. The latter two cases are mainly observed at larger impact parameters. In each incident energy, a few thousand to ten thousand events are generated, depending on the fusion cross section, proportional to the impact parameter in the given range.

The AMD calculations were performed up to times ranging from 3000 fm/c at lower energies to 1000 fm/c at higher energy side and clusterized at the end of the calculation, using a coalescence technique in phase space. The coalescence radius, corresponding to 5 fm in the coordinate space, is used at all energies. The Z, A, excitation energy, angular momentum and velocity vector of each cluster were calculated. Even after such a relatively long time, most clusters were in an excited state. In order to compare the simulated results to those of the experiments, the excited fragments were cooled down using the statistical decay code, GEMINI~\cite{Charity88}. In this calculation, the C++ version of GEMINI was used. These events are referred to as the AMD + GEMINI events hereafter, whereas the events without the GEMINI calculation are called the primary events and referred to as the AMD events.
The occurrence of the fusion reactions in the AMD + GEMINI events is defined here by the emission of the fragments with $Z > 6$ in a given event.

\begin{figure}[h]
\begin{center}
\includegraphics[scale=0.45]{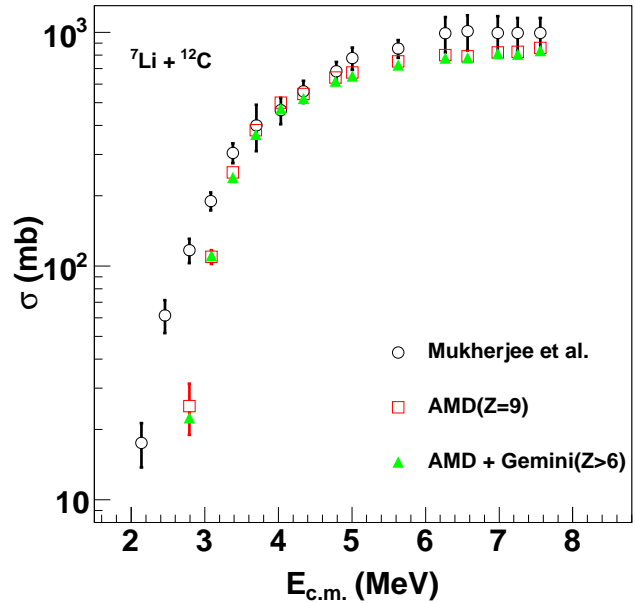}
\caption{
Fusion cross section for the $^{7}Li+^{12}C$.  Circles represent experimental results and taken from ~\cite{Mukherjee96}. Squares are the primary of AMD results filtered by $Z = 9$. Secondary values of the AMD + GEMINI events filtered by $Z > 6 $ are showed as triangles.
}
\label{fig:fig3}
\end{center}			
\end{figure}

In Fig.\ref{fig:fig3} the calculated fusion cross sections, indicated by closed triangles, are compared to those of the experiments (open circles). The experimental data are taken from ref.~\cite{Mukherjee96}. The experimental data are reproduced well within the experimental errors above $E_{cm} > 3 MeV$ in the absolute scale. The absolute cross sections predicted by the AMD simulations were calculated using the number of events generated in the given impact parameter range. At $E_{cm} \le 3 MeV$ the AMD simulation underestimated the fusion cross sections. In this energy range, the tunneling effect through the Coulomb barrier becomes important and in the present AMD formulation, this process is not incorporated. In the figure the formation cross sections of $^{19}F$ in the primary AMD events are also plotted by open square symbols. As discussed below, there are additional $20-30\%$ incomplete fusion contribution in the primary fusion process.

\begin{figure}[h]
\begin{center}
\includegraphics[scale=0.45]{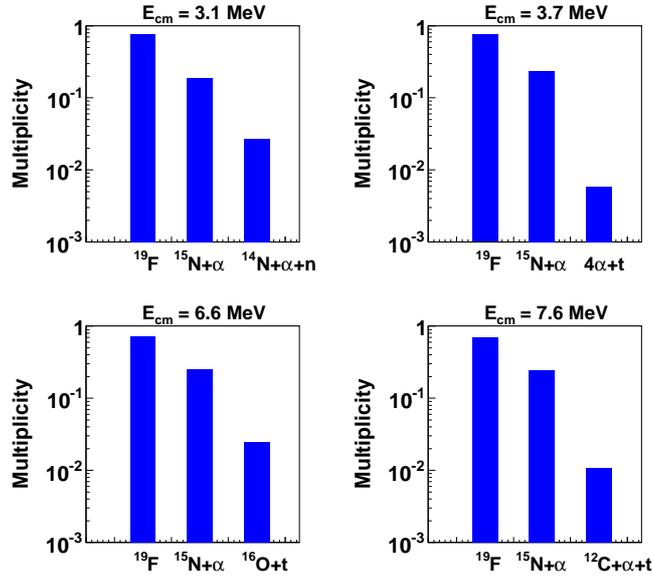}
\caption{
Primary major exit channel distribution at different incident energies.
}
\label{fig:fig4}
\end{center}			
\end{figure}

\begin{figure}[h]
\begin{center}
\includegraphics[scale=0.45]{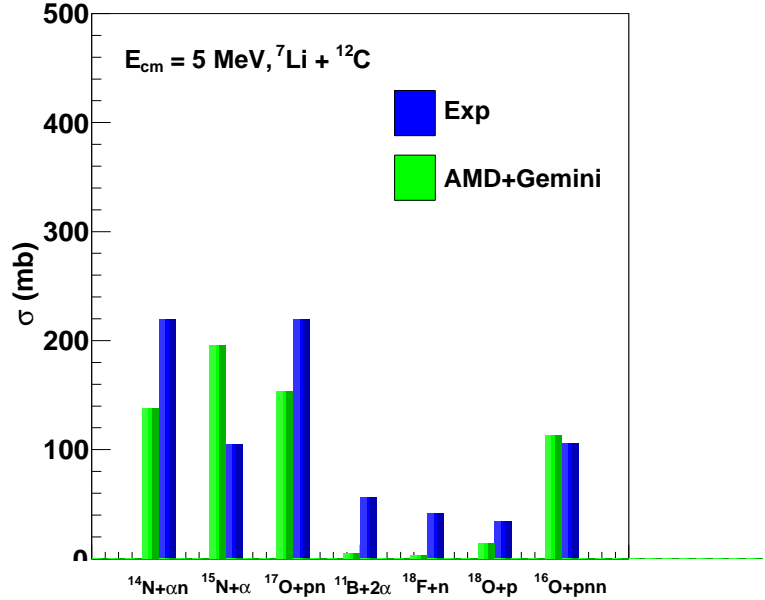}
\caption{
 Final exit channel distribution of the fusion reactions for the $^{7}Li$ + $^{12}C$.  The blue histogram indicated the experimental values. The results of AMD+GEMINI are shown by green histograms. pn and pnn channels also include d and t decays.
}
\label{fig:fig5}
\end{center}			
\end{figure}

In Fig.\ref{fig:fig4}, the fusion channel distribution at the primary stage is shown as a probability distribution. Only the top three major channels are plotted. The $^{19}F$ formation and $^{15}N+\alpha$ channel dominate the fusion reaction at all energies. Complete fusion occurs in about $80\%$ of the cases at the lower incident energies and decreases to about $70\%$ at the higher energies. The third channel contribution is from different reactions at different incident energies, but their probabilities are only a few $\%$ at most.

In Fig.\ref{fig:fig5}, the final exit channel distribution of the fusion reaction is plotted from the AMD + GEMINI events and compared with the experimental results of ref.~\cite{Mukherjee96}. The relative cross section of the major decay channels is fairly well reproduced except the $^{15}N+\alpha$ channel. The suppression of this channel in the experimental results is not yet fully understood. Further study is now underway.

\section{SUMMARY}
The fusion cross section of the $^7Li + ^{12}C$ reaction was studied using the AMD and GEMINI codes. The AMD+GEMINI simulation reproduced the experimental total fusion cross sections reasonably well at $E_{cm} > 3 MeV$ but underestimated it below that energy. The relative experimental exit channel distribution, except the $^{15}N + \alpha$ channel, was well reproduced by the AMD+GEMINI simulation.

\ack
We thank A. Ono for helpful discussions and communications. And thank him and R. Charity for letting us to use their calculation codes. One of us (R. Wada) thanks "the visiting professorship of senior international scientists" of the Chinese Academy of Sciences for the support. This work is supported by National Natural Science Foundation of China (Grant No. 11075189, 11075190 and 11005127) and Directed Program of Innovation Project of the Chinese Academy of Science (Grant No. KJCX2-YW-N44). Z.Chen thanks the "100 Person Project" of the Chinese Academy of Science. And thank the high-performance computing center of College of Physics and Information Engineering, Henan Normal University.

\end{document}